# Magnetic Circular Dichroism spectroscopy in epitaxial $La_{0.7}Sr_{0.3}MnO_3$ thin films

T. K. Nath[1] and J. R. Neal[2], G. A. Gehring[2]

[1]*Dept. of Physics and Meteorology, Indian Institute Technology of Kharagpur, 721302, W.B. India*
[2]*Dept. of Physics & Astronomy, The University of Sheffield, S3 7RH, U.K.*

*Abstract*

Magneto – optic measurements are a very powerful tool for investigating the polarization of a conduction band as a function of temperature and are used here to study the polarization of the mobile electrons in 500 Å $La_{0.7}Sr_{0.3}MnO_3$ strained thin films grown epitaxially on single crystalline (001) $LaAlO_3$ (LAO) and (001) $La_{0.3}Sr_{0.7}Al_{0.65}Ta_{0.35}O_9$ (LSAT) substrates. The magnetic circular dichroism (MCD) has been investigated in magnetic fields up to 0.5 T and over a temperature range (10 – 450 K). The MCD spectra of both the films show a peak at the band gap at around 3 eV and the peak is found to be shifted towards lower energy side with the increase of temperature. A separate polaron peak (well known in insulating samples) appears at lower energy (~1.8 eV) with the increase of temperature in all these metallic films. The rapid decrease in conduction band polarization in the film on LAO has strong implications for the use of these manganites in room temperature spintronics.

**Corresponding author: g.gehring@shef.ac.uk**

# I. INTRODUCTION

Since the discovery of negative colossal magneto-resistance (CMR) in hole doped manganese perovskite, there has been a surge of interest in manganites due to their extremely high degree of spin polarization [1,2] which is useful for potential technological applications. Band structure calculation of $La_{0.7}Sr_{0.3}MnO_3$ predicts that this system is very close to being half-metallic [3,4]. Several studies focused on determining the spin polarization, in doped lanthanum perovskites, using mainly spin-polarized tunneling measurements [1,2,5]. Proof of 100% spin polarization is still absent in manganites [6]. For doping concentrations in the range $0.2 < x < 0.5$ of these manganites, the so called Zener's "double exchange" (DE) interactions explains the paramagnetic to ferromagnetic phase transition in terms of the Mn d-electrons, namely the strong Hund's coupling between the three electrons localized in the $t_{2g}$ orbitals and the (1-x) electrons in the $e_g$ orbitals. While DE qualitatively describes the metal-insulator transitions at Tc, previous experiments [7,8] and theoretical work [9] indicated the importance of coupling between charge and the lattice, specifically the dynamic Jahn-Teller (JT) effect. Below Tc, itinerant conduction results from ferromagnetic ordering increasing the width of the $e_g$ band and suppressing the JT effect [10].

Magneto-optical Kerr effects was reported by Yamaguchi et al. [11] for hole-doped $La_{1-x}Sr_xMnO_3$ ($0 \leq x \leq 0.3$) in the photon energy range between 0.9 and 5.3 eV at room temperature. The spectral feature at ~1.2 eV was attributed to a charge transfer excitation; the feature at ~3.5 eV was attributed to electron excitations from O 2p to the majority-spin $e_g$ band ($t_{2g}^3 e_g^1 \rightarrow t_{2g}^3 e_g^2$) and to the minority-spin $t_{2g}$ band ($t_{2g}^3 e_g^1 \rightarrow t_{2g}^4 e_g^1$), respectively. Faraday spectra taken from epitaxial $La_{1-x}Ca_xMnO_3$ films [12] found a strong at ~1.5eV identified as ($t_{2g}$) → 3d ($e_g$) excitation and at 3eV identified as due to charge transfer.

In this paper we report measurements of magnetic circular dichroism (MCD) spectroscopy to study the polarization of the mobile electrons and to obtain information about spin-dependent carrier dynamics below Tc in 500 Å rf-magnetron sputtered $La_{0.7}Sr_{0.3}MnO_3$ strained thin films grown epitaxially on single crystalline (001) $LaAlO_3$ (LAO) and (001) $La_{0.3}Sr_{0.7}Al_{0.65}Ta_{0.35}O_9$ (LSAT) substrates. The MCD spectroscopy has been investigated in magnetic fields up to 0.5 T and over a temperature range (10 – 450 K). An MCD refers to a magnetically-induced difference between the optical responses to the incident light with right- and left-circular polarizations. The MCD signal depends on the transition strength, the net electron spin polarization, and the spin-orbit coupling strength. This makes the magneto-optical effects sensitive to the magnetic electron states, i.e., 3d states of Mn ions in these manganites. Since it is only dependent upon transitions at ω, it is only non zero if the crystal is absorbing. This makes very useful in determining the nature of magnetic state and any electronic transition involved. We concentrate here on the spectral region close to 1- 4.5 eV in this ferromagnetic $La_{0.7}Sr_{0.3}MnO_3$ CMR manganite. Park et al. [13] reported earlier magnetism at surface boundary ($M_{SB}$) of the half-metallic ferromagnetic $La_{0.7}Sr_{0.3}MnO_3$ thin film using temperature dependent spin-resolved photo-emission spectroscopy (SPES). They observed that the $M_{SB}$ attains a full moment at very low temperature but decays much faster than the bulk magnetization ($M_B$) upon heating. In the present investigation we have also shown the similar behavior that the maximum value of MCD peak from LSMO thin film although the MCD is sensitive to the electronic excitations in the *whole* film.

## II. EXPERIMENTAL DETAILS

The 500 Å $La_{0.7}Sr_{0.3}MnO_3$ (LSMO) films were grown epitaxially by $90^0$ off-axis rf-magnetron sputtering on different single crystalline substrates, namely (001) $LaAlO_3$ (LAO) and (001) $La_{0.3}Sr_{0.7}Al_{0.35}Ta_{0.35}O_9$ (LSAT) at 750 $^0$C from a stoichoimetric target with 200 mTorr argon-oxygen mixture [18]. The substrates were chosen for their different lattice mismatch with LSMO. In the case of LSMO film grown on LAO substrate, the lattice mismatch between them is ~ -2.3 % compressive while growing on LSAT, it is nearly lattice matched. The three-dimensional (3D) lattice parameters of the films were measured by normal and grazing incidence x-ray diffraction techniques employing four circle x-ray diffractometer with Cu $K\alpha$ radiation ($\lambda$ = 1.5418 Å). Surface morphology of both the films was studied using high resolution scanning tunneling microscopy (STM) (Digital Instrument). Temperature dependent magnetic behavior of the LSMO films were measured using superconducting quantum interference device (SQUID) magnetometry (Quantum design) in the temperature range between 5 and 400 K. Temperature dependence MCD measurements were done on both the films in transmission geometry in the temperature range of 10 – 450 K and in the magnetic field range of 0 - ±0.5 T using a specially designed MCD measurement set up. The set up is based upon the Sato method [15,16] that allows the simultaneous measurement of the two magneto-optical parameters, MCD and Faraday rotation. The set up consists of a tungsten or xenon lamp, a spectrometer (Spectro-275), a UV prism polarizer, an electromagnet, a photoelastic modulator (PEM), an analyzer and a photomultiplier tube. For low temperature measurements, the samples were mounted in a cold finger cryostat with a temperature range 10 – 450 K. An electromagnet provided a maximum magnetic field of ±0.5 T with the cryostat.

## III. RESULTS AND DISCUSSION

From the normal incidence x-ray diffraction the LSMO thin films are found to be oriented cube-on-cube type epitaxial arrangements on the substrates. The out-of-plane and in-plane lattice parameters of the 500 Å LSMO films grown on LAO substrate are (3.989 ± 0.002) Å and (3.820 ± 0.003) Å, respectively, indicating an in-plane biaxial compressive strain with $\varepsilon_{xx} = \varepsilon_{yy} = -1.3$ % and a corresponding out-of-plane tensile strain with $\varepsilon_{zz} = 3.1$ %. In the 500 Å film grown on nearly lattice matched LSAT substrate, the respective lattice parameters are (3.892 ± 0.001) Å and (3.870 ± 0.010) Å, leading to a reduced distortion with $\varepsilon_{xx} = \varepsilon_{yy} = 0$ and $\varepsilon_{zz} = 0.57$ % when compared to its LAO counterpart. The substrate induced lattice strain can be decomposed into bulk strain, $\varepsilon_B$ and a volume-preserving Jahn-Teller (JT) strain, $\varepsilon_{JT}$ [18,21]. The perovskite unit cell volume ($V_F$), bulk strain ($\varepsilon_B$), JT strain ($\varepsilon_{JT}$) for the 500 Å LSMO film grown on LAO substrates are 58.209 Å$^3$, +0.50 % and +3.59%, respectively. In the case of the film grown on nearly lattice matched LSAT substrate, the corresponding values are found to be 58.290 Å$^3$, +0.57 % and +0.46 %, respectively.

Figure 1 (a) shows the typical raw data of MCD spectra of 500 Å LSMO film grown on (001) LAO substrate at different temperatures in a magnetic field of 0.5 T applied along [001]. The MCD spectra of the LSMO film show a strong peak at the band gap at around 3.3 eV at the lowest temperature of 10 K and the peak is found to be shifted towards lower energy side with the increase of temperature. The MCD signal at ~3.3eV weakens rapidly with temperature and a separate low energy broad peak appears at around 1.8 eV photon energy in the MCD spectra recorded above 250 K. Hysteresis loops measured using MCD show coercive fields that agree with those measured by SQUID magnetometry [17].

Figure 2 shows similar MCD spectra for a 500 Å LSMO film grown on nearly lattice matched (001) LSAT substrate. The strong band gap signal is seen again and the low energy broad peak at ~1.7 eV is observable above 300 K. These films show a smaller coercive field than for those grown on LAO.

Finally we focus on the temperature dependence of the MCD, which may be assumed to be proportional to the polarization of the conduction electrons of the LSMO films. The comparative study of temperature dependence of the normalized magnetization [M(T)/M(10 K)] obtained from SQUID measurement with the temperature dependence normalized MCD signal at the strong peak observed at 3.3 eV photon energy $[MCD(T)/MCD(10\ K)]_{peak\ (3.3\ eV)}$ for the LSMO films on LAO and LSAT are shown in Figure 3. The two films show strikingly different behavior. The MCD intensity follows the magnetization for the film grown on LSAT but shows a remarkable deviation for the films grown on LAO.

## IV. CONCLUSIONS

We have shown the magneto-optical studies are a very powerful technique to study manganite films. The magnitude of the signal is proportional to the polarization of the band electrons. It is this polarization that is needed in spintronics. Our results show that LSAT is a very promising substrate for LSMO films for spintronic applications. In addition we have seen the emergence of the polaronic peak expected in the insulating phase below the ferromagnetic transition temperature in both of our films.

## V. ACKNOWLEDGEMENT

One of us (T.K.N) would like to acknowledge Royal Society, London for financial assistance under the scheme on grants for overseas scientist for short term visit to carry out the experiments in the University of Sheffield, U.K. We also thank EPSRC for support of JRN and these experimental facilities.

# Figure Captions

Figure 1. MCD measurements of 500 Å epitaxial LSMO films grown on (001) LaAlO$_3$ substrate at T = 10, 50, 100, 150, 200, 250, 300 and 350 K in a magnetic field of 0.5 T applied along [001] direction of the film.

Figure 2. MCD measurements of 500 Å epitaxial LSMO films grown on (001) LSAT substrate at T = 10, 50, 100, 150, 200, 250, 300 and 350 K in a magnetic field of 0.5 T applied along [001] direction of the film.

Figure 3. (a) Temperature dependence of M(T)/M(10 K) and MCD(T)/MCD(10K) for 500 Å epitaxial LSMO films grown on (a) LAO and (b) LSAT single crystalline substrates. It shows that the temperature dependence of surface boundary magnetization (MCD) is different from the bulk magnetization.

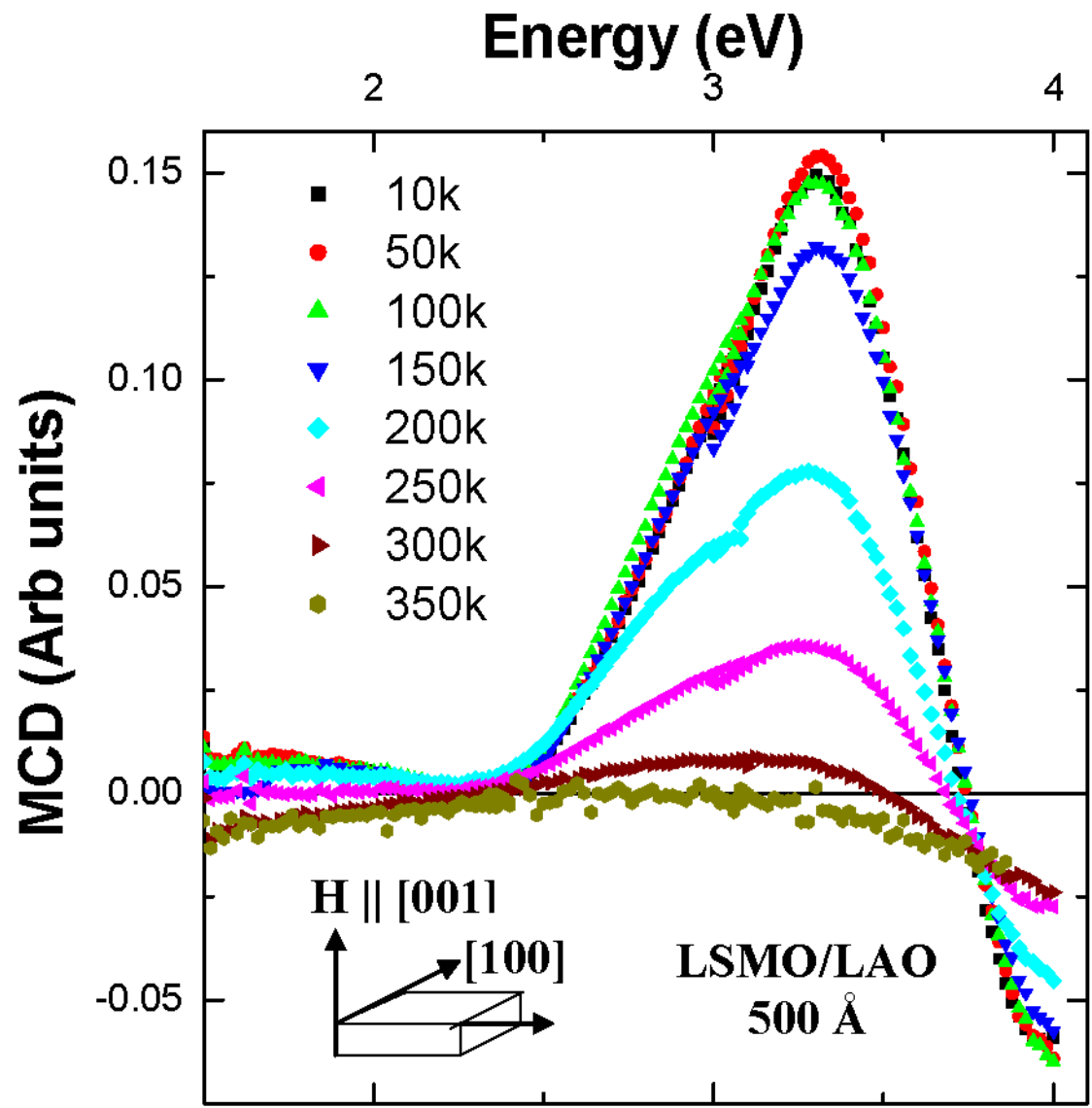

Figure 1: G. Gehring et al.

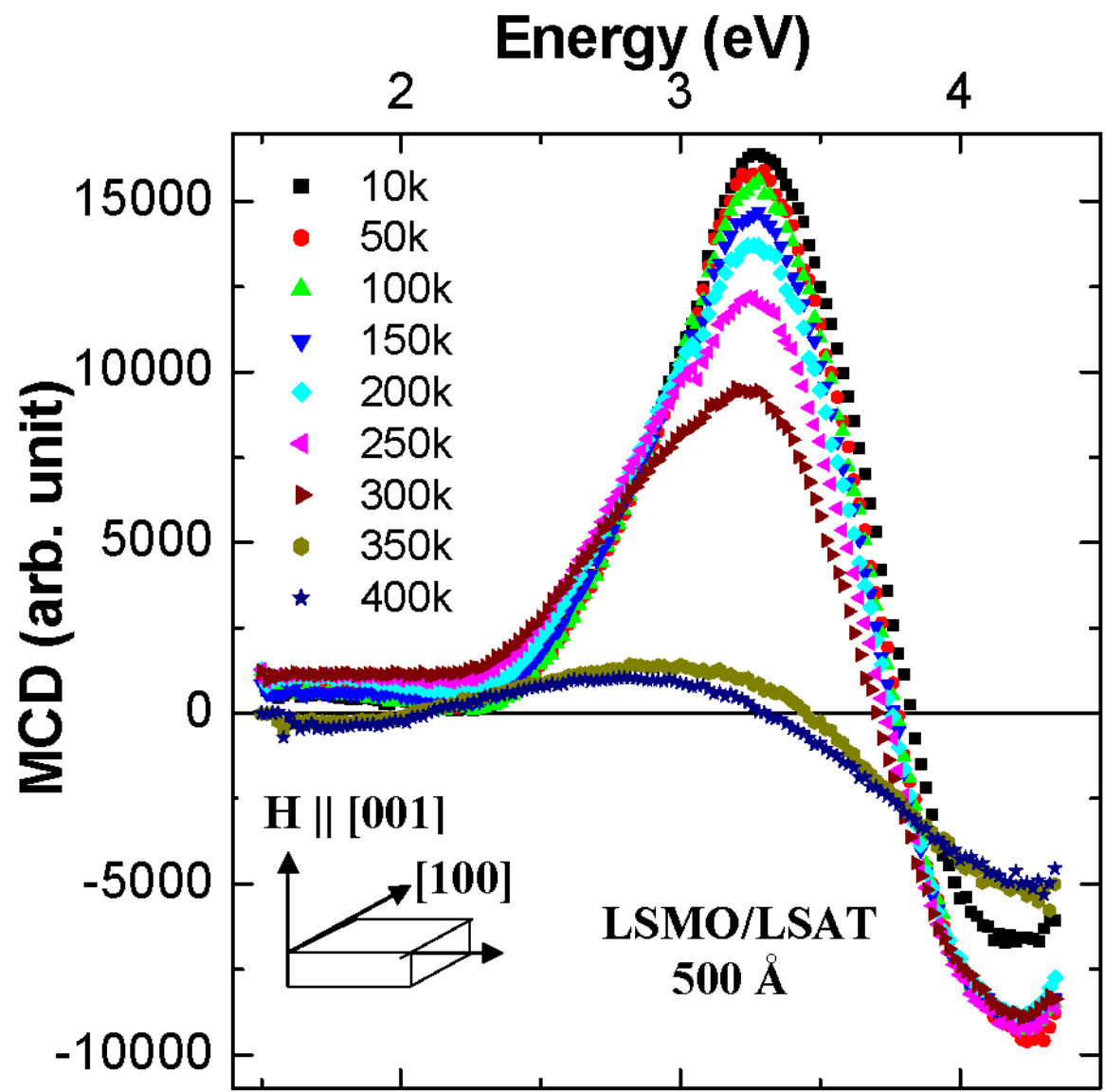

Figure 2: G. Gehring et al.

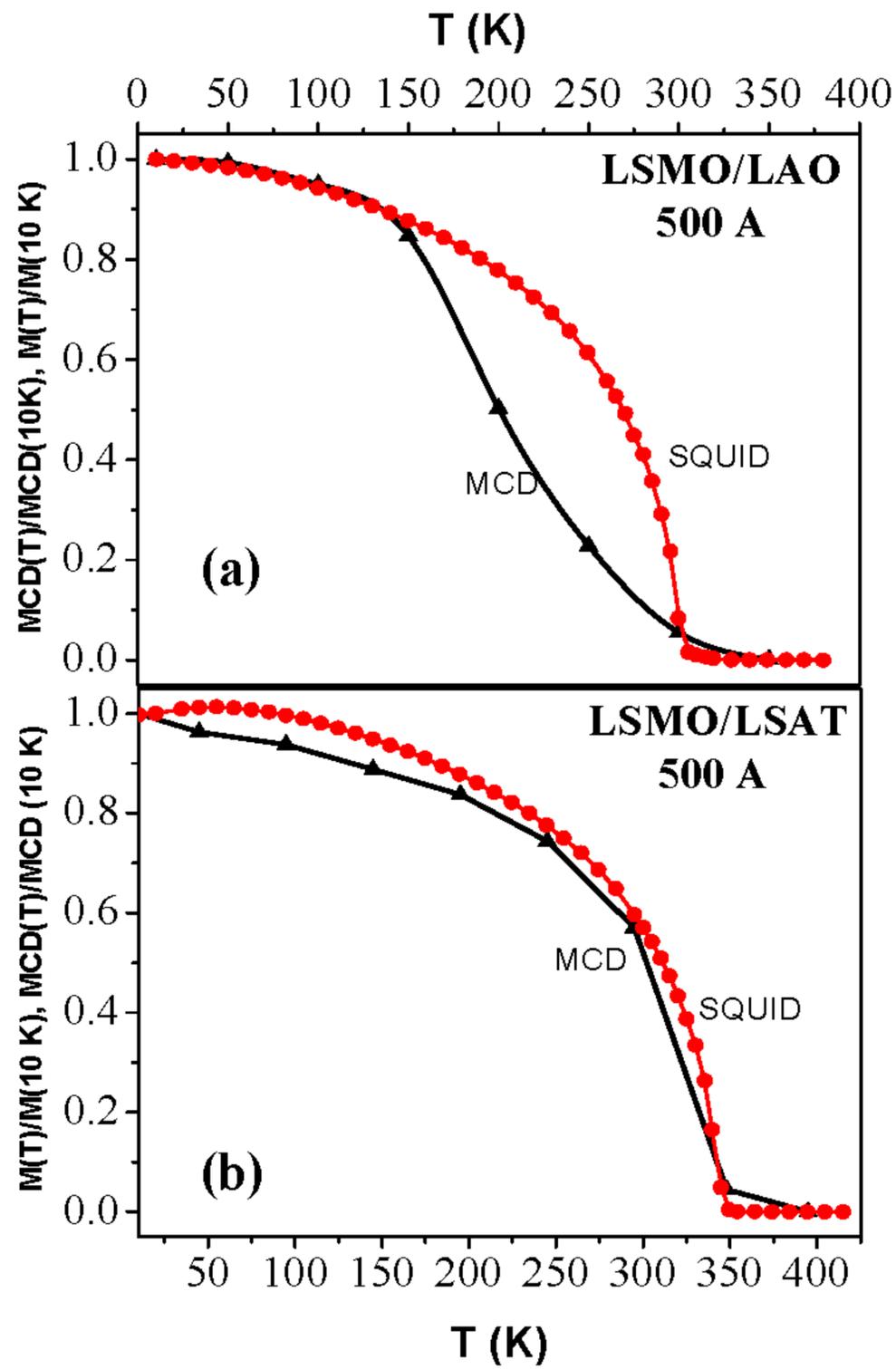

Figure 3: G. Gehring et al.

1